\documentclass{elsart}
\usepackage{graphicx,amssymb}
\journal{Physica A}

\begin{document}
\begin{frontmatter}
\title{Cross-Correlation Dynamics in Financial Time Series}
\author[DCU]{T. Conlon},
\ead{tconlon@computing.dcu.ie}
\author[DCU]{H.J. Ruskin},
\ead{hruskin@computing.dcu.ie}
\author[DCU]{M. Crane},
\ead{mcrane@computing.dcu.ie}
\address[DCU]{Dublin City University, Glasnevin, Dublin 9, Ireland}

\begin{abstract}
The dynamics of the equal-time cross-correlation matrix of multivariate financial time series is explored by examination of the \textit{eigenvalue spectrum} over sliding time windows.  Empirical results for the S\&P 500 and the Dow Jones Euro Stoxx 50 indices reveal that the dynamics of the small eigenvalues of the cross-correlation matrix, over these time windows, oppose those of the largest eigenvalue.  This behaviour is shown to be independent of the size of the time window and the number of stocks examined.  

A basic one-factor model is then proposed, which captures the main dynamical features of the eigenvalue spectrum of the empirical data.  Through the addition of perturbations to the one-factor model, (leading to a `market plus sectors' model), additional sectoral features are added, resulting in an Inverse Participation Ratio comparable to that found for empirical data.  By partitioning the eigenvalue time series, we then show that negative index returns, (\emph{drawdowns}), are associated with periods where the largest eigenvalue is greatest, while positive index returns, (\emph{drawups}), are associated with periods where the largest eigenvalue is smallest. The study of correlation dynamics provides some insight on the collective behaviour of traders with varying strategies.

\end{abstract}
\begin{keyword}
Correlation Matrix \sep Eigenspectrum Analysis \sep Econophysics
\PACS {89.65.Gh}   \sep
      {89.65.-s}	 \sep
      {89.75.-k}
\end{keyword}
\end{frontmatter}

\section{Introduction}
In recent years, the analysis of the equal-time cross-correlation matrix for a variety of multivariate data sets such as financial data, \cite{Laloux_1999,Plerou_1999,Laloux_2000,Plerou_2000,Gopikrishnan_2000,Utsuki_2004,Bouchaud_book,Wilcox_2004,Sharifi_2004,Conlon_2007,Conlon_2008,Podobnik_2008}, electroencephalographic (EEG) recordings, \cite{Schindler_2007,Schindler_2007_b}, magnetoencephalographic (MEG) recordings, \cite{Kwapien_2000}, and others, has been studied extensively.  Other authors have investigated the relationship between stock price changes and liquidity or trading volume, \cite{Ying_1966, Karpoff_1987,LeBaron_1999}.  In particular, Random Matrix Theory, (RMT), has been applied to filter the relevant information from the statistical fluctuations, inherent in empirical cross-correlation matrices, for various financial time series, \cite{Laloux_1999,Plerou_1999,Laloux_2000,Plerou_2000,Gopikrishnan_2000,Utsuki_2004,Bouchaud_book,Wilcox_2004,Sharifi_2004,Conlon_2007}.  By comparing the eigenvalue spectrum of the correlation matrix to the analytical results, obtained for random matrix ensembles, significant deviations from RMT eigenvalue predictions provide genuine information about the correlation structure of the system.  This information has been used to reduce the difference between predicted and realised risk of different portfolios.

Several authors have suggested recently that some real correlation information may be hidden in the RMT defined random part of the eigenvalue spectrum.  A technique, involving the use of power mapping to identify and estimate the noise in financial correlation matrices, \cite{Guhr_2003}, allows the suppression of those eigenvalues, associated with the noise, in order to reveal different correlation structures buried underneath.  The relationship, between the eigenvalue density $c$ of the true correlation matrix, and that of the empirical correlation matrix $C$, was derived to show that correlations can be measured in the random part of the spectrum, \cite{Burda_2004_a, Burda_2004_b}. A Kolmogorov test was applied to demonstrate that the bulk of the spectrum is not in the Wishart RMT class, \cite{Malevergne}, while the existence of factors, such as an overall market effect, firm size and industry type, is due to collective influence of the assets.  More evidence that the RMT fit is not perfect was provided, \cite{Kwapien_2006}, where it was shown that the dispersion of ``noise'' eigenvalues is inflated, indicating that the bulk of the eigenvalue spectrum contains correlations masked by measurement noise.

The behaviour of the largest eigenvalue of a cross-correlation matrix for small windows of time, has been studied, \cite{Drozdz_2000}, for the DAX and Dow Jones Industrial average Indices (DJIA).  Evidence of a time-dependence between `drawdowns' (`draw-ups') and an increase (decrease) in the largest eigenvalue was obtained, resulting in an increase of the \textit{information entropy}\footnote{In information theory, the Shannon entropy or information entropy is a measure of the uncertainty associated with a random variable.}of the system. Similar techniques were used, \cite{Drozdz_2001}, to investigate the dynamics between the stocks of two different markets (DAX and DJIA).  In this case, two distinct eigenvalues of the cross-correlation matrix emerged, corresponding to each of the markets.  By adjusting for time-zone delays, the two eigenvalues were then shown to coincide, implying that one market leads the dynamics in the other.

Equal-time cross-correlation matrices have been used, \cite{Muller_2005}, to characterise dynamical changes in nonstationary multivariate time-series.  It was specifically noted that, for increased synchronisation of $k$ series within an $M-$dimensional multivariate time series, a repulsion between eigenstates of the correlation matrix results, in which $k$ levels participate.  Through the use of artifically-created time series with pre-defined correlation dynamics, it was demonstrated that there exist situations, where the relative change in eigenvalues from the lower edge of the spectrum is greater than that for the large eigenvalues, implying that information drawn from the smaller eigenvalues is highly relevant.  

The technique was subsequently applied to the eigenvalue spectrum of the equal time cross-correlation matrix of multivariate Epileptic Seizure time series and information on the correlation dynamics was found to be visible in \emph{both} the lower and upper eigenstates.  Further studies of the equal-time correlation matrix of EEG signals, \cite{Schindler_2007}, investigated temporal dynamics of focal onset epileptic seizures\footnote{A partial or focal onset seizure occurs when the discharge starts in one area of the brain and then spreads over other areas.} and showed that zero-lag correlations between multichannel EEG signals tend to decrease during the first half of a seizure and increase gradually before the seizure ends.  A further extension (to the case of \textit{Status Epilepticus}, \cite{Schindler_2007_b}), used the equal-time correlation matrix to assess neuronal synchronisation prior to seizure termination.  

Examples have demonstrated, \cite{Muller_2006_b}, that information about cross correlations can be found in the RMT bulk of eigenvalues and that the information extracted at the \emph{lower} edge is statistically \emph{more significant} than that extracted from the larger eigenvalues.  The authors introduced a method of unfolding the eigenvalue level density, through the normalisation of each of the level distances by its ensemble average, and used this to calculate the corresponding individual nearest-neighbour distance.  Those parts of the spectrum, dominated by noise, could be distinguished from those containing information about correlations.  Application of this technique to multichannel EEG data showed the smallest eigenvalues to be more sensitive to detection of subtle changes in the brain dynamics than the largest. 

In this paper, we examine eigenvalue dynamics of the cross-correlation matrix for multivariate financial data with a view to characterising market behaviour.  Methods are reviewed in Section~\ref{methods}, the data described in Section~\ref{Data} and in Section~\ref{Eresults} we look at the results obtained both for the empirical correlation matrix and the model correlation matrices described.

\section{Methods}
\label{methods}
\subsection{Empirical Dynamics}
\label{Emp_Dyn}
The equal-time cross-correlation matrix, between time series of equity returns, is calculated using a sliding window where the number of assets, $N$, is smaller than the window size $T$.  Given returns $G_{i} \left(t\right)$, $i = 1, \ldots,N$, of a collection of equities, we define a normalised return, within each window, in order to standardise the different equity volatilities. We normalise $G_{i}$ with respect to its standard deviation $\sigma_{i}$ as follows:
\begin{equation}
g_{i}\left(t\right) = \frac{G_{i} \left(t\right) - \widehat{G_{i} \left(t\right)}}{\sigma_{i}}
\end{equation}
Where $\sigma_{i}$ is the standard deviation of $G_{i}$ for assets $i = 1, \ldots,N$ and $\widehat{G_{i}}$ is the time average of $G_{i}$ over a time window of size $T$.

Then, the equal-time cross-correlation matrix is expressed in terms of $g_{i}\left(t\right)$
\begin{equation}
C_{ij} \equiv \left\langle g_{i}\left(t\right) g_{j}\left(t\right) \right\rangle
\label{cross_corr}
\end{equation}
The elements of $C_{ij}$ are limited to the domain $-1 \leq C_{ij} \leq 1$, where $C_{ij} = 1$ defines perfect positive correlation, $C_{ij} = -1$ corresponds to perfect negative correlation and $C_{ij} = 0$ corresponds to no correlation. In matrix notation, the correlation matrix can be expressed as
\begin{equation}
\mathbf{C} = \frac{1}{T} \mathbf{GG}^{\tau}
\end{equation}
Where $\bf{G}$ is an $N\times T$ matrix with elements $g_{it}$.

The eigenvalues $\mathbf{\lambda}_{i}$ and eigenvectors $\mathbf{\hat{v}}_{i}$ of the correlation matrix $\mathbf{C}$ are found from the following
\begin{equation}
\mathbf{C} \mathbf{\hat{v}}_{i} = \mathbf{\lambda}_{i} \mathbf{\hat{v}}_{i}
\end{equation}
The eigenvalues are then ordered according to size, such that $\mathbf{\lambda}_{1} \leq \mathbf{\lambda}_{2}\leq \ldots \leq \mathbf{\lambda}_{N}$.  The sum of the diagonal elements of a matrix, (the Trace), must always remain constant under linear transformation. Thus, the sum of the eigenvalues must always equal the Trace of the original correlation matrix.  Hence, if some eigenvalues increase then others must decrease, to compensate, and vice versa (\textit{Eigenvalue Repulsion}).  

There are two limiting cases for the distribution of the eigenvalues \cite{Schindler_2007,Muller_2005}. When all of the time series are perfectly correlated, $C_{i} \approx 1$, the largest eigenvalue is maximised with a value equal to $N$, while for time series consisting of random numbers with average correlation $C_{i} \approx 0$, the corresponding eigenvalues are distributed around $1$, (where any deviation is due to spurious random correlations). 

For cases between these two extremes, the eigenvalues at the lower end of the spectrum can be much smaller than $\lambda_{max}$.  To study the dynamics of each of the eigenvalues using a sliding window, we normalise each eigenvalue in time using
\begin{equation}
{\tilde{\lambda}}_{i}(t) = \frac{\left(\mathbf{\lambda}_{i}(t) - {\widehat{\lambda_{i}(\tau)}}\right)}{\sigma^{\lambda_{i}\left(\tau\right)}}
\label{normalise}
\end{equation}
where $\mathbf{\widehat{\lambda}}$ and $\sigma^{\lambda}$ are the mean and standard deviation of the eigenvalue $i$ over a particular reference period, $\tau$.  This normalisation allows us to visually compare eigenvalues at both ends of the spectrum, even if their magnitudes are significantly different.  The reference period, used to calculate mean and standard deviation of the eigenvalue spectrum, can be chosen to be a low volatility sub-period, (which helps to enhance the visibility of high volatility periods), or the full time period studied.

\subsection{One-factor Model}
\label{OneFModel}
In the \emph{one-factor model} of stock returns, we assume a \emph{global correlation} with the cross-correlation between all stocks the same, $\rho_{0}$.  The spectrum of the associated correlation matrix consists of only two values, a large eigenvalue of order $(N-1)\rho_{0} + 1$, associated with the market, and an $(N-1)-$fold degenerate eigenvalue of size $1-\rho_{0}<1$.  Any deviation from these values is due to the finite length of time series used to calculate the correlations.  In the limit $N\rightarrow\infty$ (even for \emph{small correlation}, i.e. $\rho\rightarrow0$) a large eigenvalue appears, which is associated with the eigenvector $v_{1} = \left(\frac{1}{\sqrt{N}}\right)\left(1,1,1\ldots1\right)$, and which dominates the correlation structure of the system.

\subsection{`Market plus sectors' model}
\label{MultiFModel}
To expand the above to a `market plus sectors' model, we perturb a number of pairs $N$ of the correlations $\rho_{0} + \rho_{n}$, where $-1-\rho_{0}\leq\rho_{n}\leq1-\rho_{0}$.  Additionally, we impose a constraint $\displaystyle\sum_{N}\rho_n = 0$, ensuring that the average correlation of the system remains equal to $\rho_{0}$.  These perturbations allow us to introduce \textit{groups of stocks} with similar correlations, (corresponding to Market Sectors).

Using the correlation matrix from the ``one-factor model'' and the ``market plus sectors model'', we can construct correlated time series using the Cholesky decomposition $A$ of a correlation matrix $C = {AA}^{\tau}$.  We can then generate finite correlated time series of length $T$,
\begin{equation}
x_{it} = \sum_{j} A_{ij} y_{jt}  \quad \quad t = 1,\dots,T
\label{chol}
\end{equation}
where $y_{jt}$ is a random Gaussian variable with mean zero and variance $1$ at time $t$.   Using Eqn.~\ref{cross_corr} we can then construct a correlation matrix using the simulated time series.  The finite size of the time series introduces `noise' into the system and hence empirical correlations will vary from sample to sample.  This `noise' could be reduced through the use of longer simulated time series or through averaging over a large number of time series. 

In order to compare the eigenvectors from each of the Model Correlation matrices to that constructed from the equity returns time series, we use the Inverse Participation Ratio (IPR) \cite{Plerou_2000,Noh}.  The IPR allows quantification of the number of components that participate significantly in each eigenvector and tells us more about the level and nature of deviation from RMT. The IPR of the eigenvector $u^{k}$ is given by
$ I^{k} \equiv \sum^{N}_{l = 1} \left( u^{k}_{l}\right)^{4} $ and allows us to compute the inverse of the number of eigenvector components that contribute significantly to each eigenvector.

\section{Data}
\label{Data}
In order to study the dynamics of the empirical correlation matrix over time, we analyse two different data sets.  The first data set comprises the $384$ equities of the Standard \& Poors (S\&P) 500 where full price data is available from January $1996$ to August $2007$ resulting in 2938 daily returns.  The S\&P 500 is an index consisting of 500 large capitalisation equities, which are predominantly from the US. In order to demonstrate that our results are not market specific, however, we examine a second data set, made up of the 49 equities of the Dow Jones Euro Stoxx 50 where full price data is available from January $2001$ to August $2007$ resulting in $1619$ daily returns.  The Dow Jones Euro Stoxx 50 is a stock index of Eurozone equities designed to provide a blue-chip representation of supersector leaders in the Eurozone.

\section{Results}
\label{Eresults}
\begin{figure}[htbp!]
\begin{center}
\includegraphics[height=100mm,width=130mm]{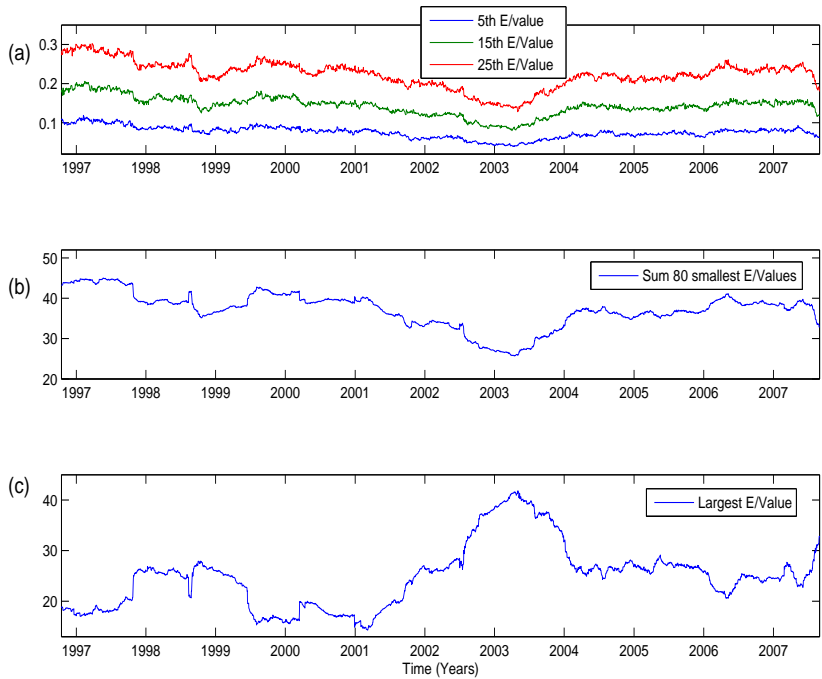}
\caption{Time Evolution of (a) Three small eigenvalues (b) Sum of the 80 smallest eigenvalues (c) The largest eigenvalue}
\label{S&Peigs}
\end{center}
\end{figure}

We analyse the eigenvalue dynamics of the correlation matrix of a subset of 100 S\&P equities, chosen randomly, using a sliding window of 200 days.  This subsector was chosen such that $Q = \frac{T}{N} = 2$, thus ensuring that the data would be close to non-stationary in each sliding window. Figure \ref{S&Peigs}(a) shows broadly similar sample dynamics from the $5^{th}$, $15^{th}$ and $25^{th}$ largest eigenvalues over each of these sliding windows.  The sum of the 80 smallest eigenvalues are shown in Figure \ref{S&Peigs}(b), while the dynamics of the largest eigenvalue is displayed in Figure \ref{S&Peigs}(c).  The repulsion between the largest eigenvalue and the small eigenvalues is evident here, (comparing \ref{S&Peigs}(b) and \ref{S&Peigs}(c)), with the dynamics of the small eigenvalues contrary to those of the largest eigenvalue.  As noted earlier (Section~\ref{Emp_Dyn}), this is a consequence of the fact that the trace of the correlation matrix must remain constant under transformations and any change in the largest eigenvalue must be reflected by a change in one or more of the other eigenvalues.  Similar results were obtained for different subsets of the S\&P and also for the members of the Dow Jones Euro Stoxx 50.

\subsection{Normalised Eigenvalue Dynamics}
\begin{figure}[htbp!]
\begin{center}
\includegraphics[height=120mm,width=130mm]{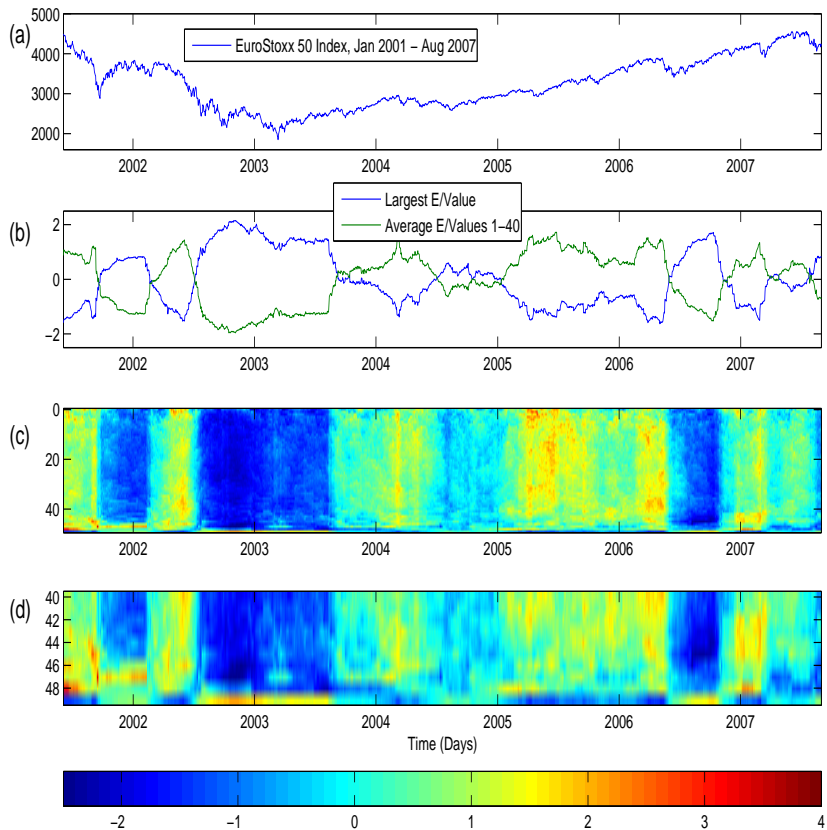}
\caption{(a) EuroStoxx 50 Index, Jan 2001 - Aug 2007 (b) Normalised Largest Eigenvalue vs. Average of 40 smallest normalised eigenvalues for EuroStoxx 50 (c) All Normalised Eigenvalues (d) The 9 Largest Normalised Eigenvalues}
\label{SX5E}
\end{center}
\end{figure}

Using normalised eigenvalues as described above, (Eqn. \ref{normalise}), we performed a number of experiments to investigate the dynamics of a set of small eigenvalues versus the largest eigenvalue.  The various experiments are described below:

\begin{enumerate}
\item
As in Section~\ref{Eresults}, the dynamics for the same subset of 100 equities are analysed using a sliding window of 200 days.  The normalisation is carried out using the mean and standard deviation of each of the eigenvalues over the entire time-period. Figure \ref{S&PeigsNorm}(a) shows the value of the S\&P index from $1997$ to mid$-2007$.  

The normalised largest eigenvalue is shown in Figure \ref{S&PeigsNorm}(b) together with the average of the 80 normalised small eigenvalues.  The compensatory dynamics mentioned earlier are shown more clearly here, with the largest and average of the smallest 80 eigenvalues having opposite movements.  The normalised eigenvalues for the entire eigenvalue spectrum are shown in Figure \ref{S&PeigsNorm}(c), where the colour indicates the number of standard deviations from the time average for each of the eigenvalues over time.  As shown, there is very little to differentiate the dynamics of the $80-90$ or so smallest eigenvalues.  In contrast, the behaviour of the largest eigenvalue is clearly opposite to that of the smaller eigenvalues.  However, from the $90^{th}$ and subsequent eigenvalue there is a marked change in the behaviour, (Figure \ref{S&PeigsNorm}(d)), and the eigenvalue dynamics are distinctly different.  This may correspond to the area outside the ``Random Bulk'' in RMT.  Similar to \cite{Drozdz_2000,Drozdz_2001}, we also find evidence of an increase/decrease in the largest eigenvalue with respect to `drawdowns'/`draw-ups'.  Additionally, we find the highlighted \textit{compensatory dynamics} of the small eigenvalues.  These results were tested for various time windows and normalisation periods, with smaller windows and normalisation periods found to capture and emphasise additional features. 

\begin{figure}[htbp!]
\begin{center}
\includegraphics[height=120mm,width=130mm]{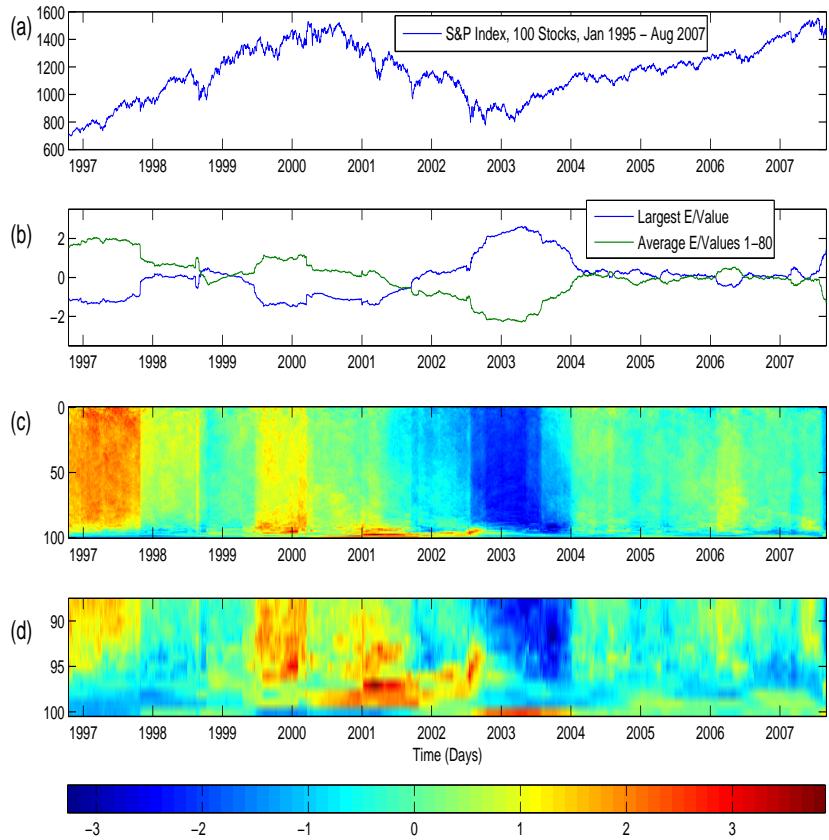}
\caption{(a) S\&P Index (b) Normalised Largest Eigenvalue vs. Average of 80 smallest normalised eigenvalues (c) All Normalised Eigenvalues (d) Largest 12 Normalised Eigenvalues}
\label{S&PeigsNorm}
\end{center}
\end{figure}

\item
To demonstrate the above result for a \emph{different level of granularity}, we chose $50$ equities randomly with a time window of $500$ days, giving $Q = \frac{T}{N} = 10$.  The results obtained, (Figure \ref{S&Peigs50}), are in keeping with those for $Q = 2$ earlier, with a broad-band increase (decrease) of the $40$ smallest eigenvalues concurrent to a decrease (increase) of the largest eigenvalue.

\begin{figure}[htbp!]
\begin{center}
\includegraphics[height=70mm,width=130mm]{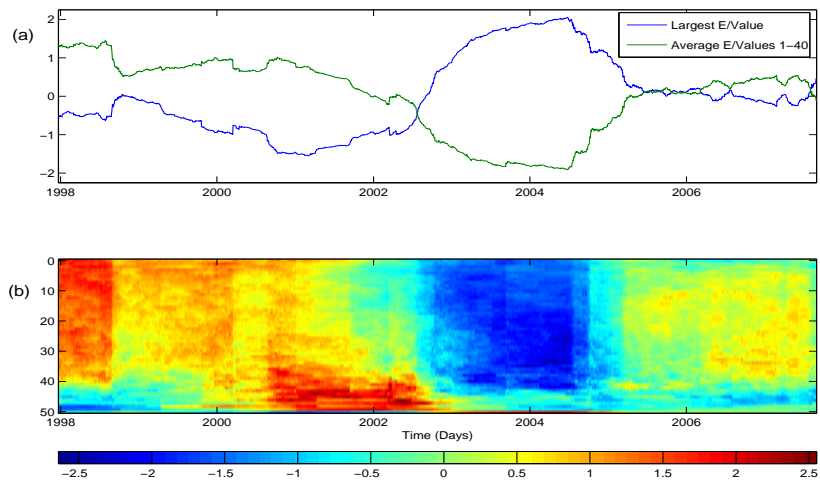}
\caption{(a) Normalised Largest Eigenvalue vs. Average of 40 smallest normalised eigenvalues (b) All Normalised Eigenvalues}
\label{S&Peigs50}
\end{center}
\end{figure}

\item
The previous examples used random subsets of the S\&P universe in order to keep $Q= \frac{T}{N}$ as large as possible.  To demonstrate that the above results were not sampling artifacts, we also looked at the full sample of 384 equities, (that survived the entire $11$ year period), with a time window of $500$ days ($Q = 1.30$).  The results, as shown in Figure \ref{S&Peigs384}, are similar to those above, with the majority of the small eigenvalues compensating for changes in the large eigenvalue.   As indicated previously, however, there is a small band of large eigenvalues, for which behaviour is different to that of both the band of small eigenvalues and the largest eigenvalue.

\begin{figure}[htbp!]
\begin{center}
\includegraphics[height=70mm,width=130mm]{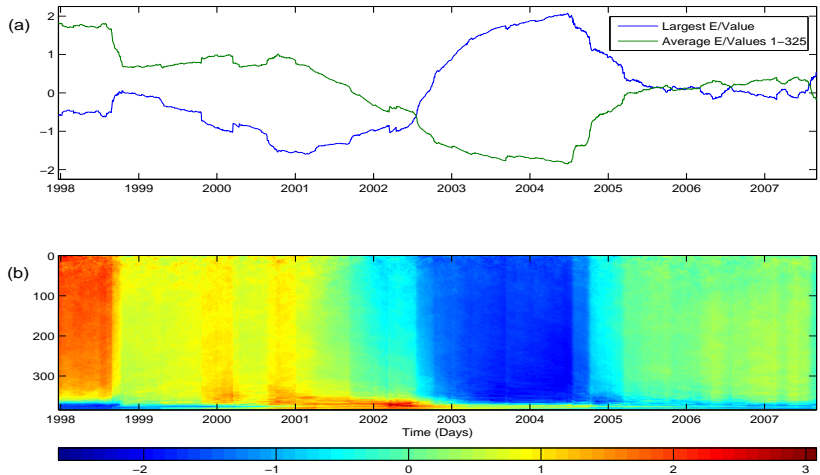}
\caption{(a) Normalised Largest Eigenvalue vs. Average of 325 smallest normalised eigenvalues (b) All Normalised Eigenvalues}
\label{S&Peigs384}
\end{center}
\end{figure}

\item
All examples discussed so far have focused on the universe of equities from the S\&P $500$ that have survived since $1997$.  To ensure that the results obtained were not exclusive to the S\&P $500$, we also applied the same technique to the $49$ equities of the EuroStoxx 50 index that survived from January $2001$ to August $2007$.  The sliding window used was $200$ days, such that $Q = 4.082$.  The results, (Figure \ref{SX5E}), were again similar to before, with a wide band of small eigenvalues ``responding to'' movements in the largest eigenvalues.  In this case, the band of deviating large eigenvalues (ie. those which correspond to the area outside the ``Random Bulk'' in RMT), (Figure \ref{SX5E}(d)), is not as marked as in the previous example. This effectively implies that equities in this index are dominated by ``the Market''.

\end{enumerate}

\subsection{Model Correlation Matrix}
\label{corr_model}
The results, described, demonstrate that the time dependent dynamics of the small eigenvalues of the correlation matrix of stock returns move counter to those of the largest eigenvalue.  Here, we show how a simple one-factor model, Section~\ref{OneFModel}, of the correlation structure reproduces much of this behaviour.  Furthermore, we show how additional features can be captured by including perturbations in this model, essentially a \textit{``market plus sectors''} model, Section~\ref{MultiFModel}, \cite{Malevergne,Noh,Papp}.  

\begin{figure}[htbp!]
\begin{center}
\includegraphics[height=95mm,width=135mm]{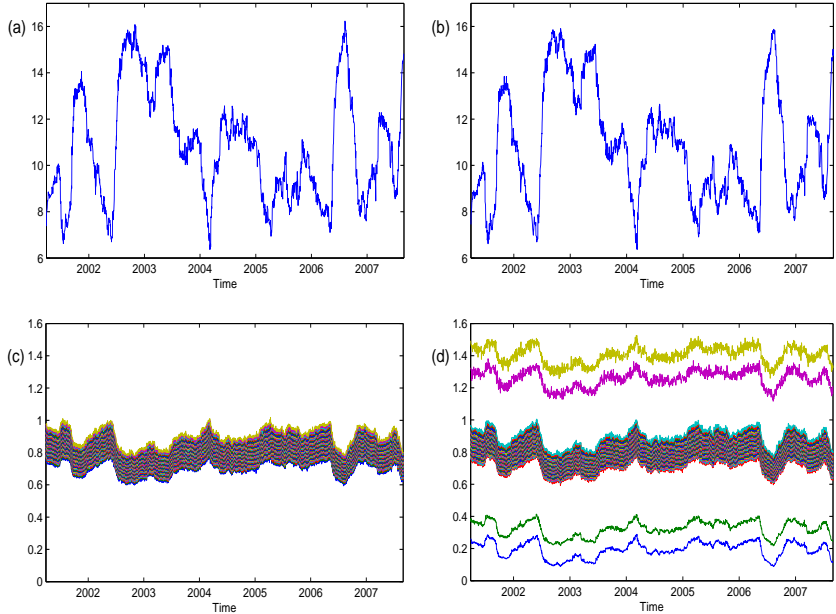}
\caption{(a) Largest Eigenavalue, one factor model (b) Largest Eigenavalue, Market plus sectors model (c) Eigenvalues 1-48, one factor model (d) Eigenvalues 1-48, Market plus sectors model}
\label{modelEuroStoxx}
\end{center}
\end{figure}

In order to compare the empirical results, Section~\ref{Eresults}, to those of the single factor model, we first constructed a correlation matrix where each non-diagonal element was equal to the average correlation of the empirical matrix in each sliding window.  We then calculated the eigenvalues of this matrix over each sliding window and normalised these as before, (Section \ref{methods}).  The dynamics of the largest eigenvalue for the single-factor model are displayed in Figure \ref{modelEuroStoxx}(a) for the EuroStoxx 50 index using a sliding window of $200$ days.  As can be seen, the main features of the dynamics are in agreement with those of Figure~\ref{SX5E} for the empirical data. The dynamics of the remaining eigenvalues, shown in figure \ref{modelEuroStoxx}(c), are found to be within a narrow range, with any change in time due to compensation for change in the largest eigenvalue.

For the `market plus sectors' model, we introducted perturbations with two groups of $5$ stocks having correlation $\rho_{0}-0.15$ and $\rho_{0}+0.15$, with the average correlation at each time window remaining the same. 
The largest eigenvalue dynamics for the `market plus sectors' model is shown in \ref{modelEuroStoxx}(b), with the dynamics almost identical to those for the largest eigenvalue of the `one-factor' model, (any differences are due to random fluctuations induced in the Cholesky decomposition, Section~\ref{MultiFModel}).  However, looking at the remaining eigenvalues, \ref{modelEuroStoxx}(d), we see that a number are found to be deviate significantly from the bulk.  These deviations are due to the additional sectorial information included in the market plus sectors model.  This agrees with previous results, \cite{Plerou_2000, Utsuki_2004}, where small eigenvalues were found to deviate from the random bulk, in addition to large deviating eigenvalues containing sectorial information.

\begin{figure}[htbp!]
\begin{center}
\includegraphics[height=70mm,width=130mm]{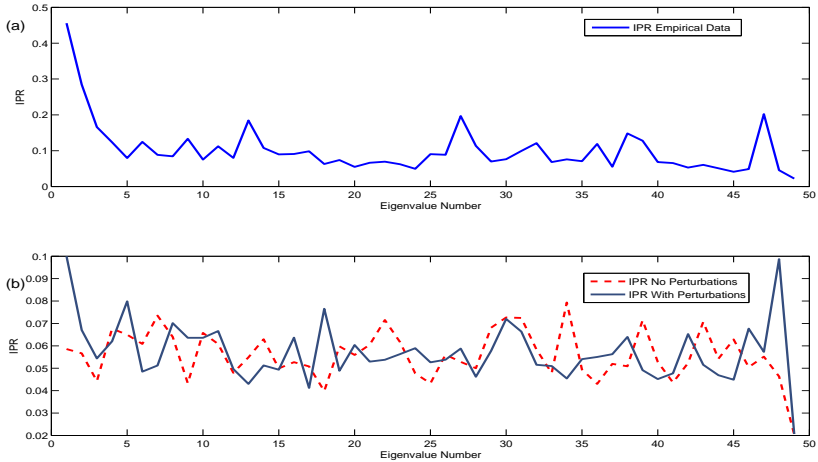}
\caption{(a) IPR Empirical Data (b) IPR Simulated Data}
\label{IPR}
\end{center}
\end{figure}

To examine properties of the eigenvector components themselves, we use the Inverse Participation Ratio (IPR).  Figure~\ref{IPR}(a), displays the IPR found using the empirical data from the Eurostoxx $50$.  This has similar characteristics to those found for different indices, \cite{Plerou_2000}, with the IPR for the largest eigenvalue much smaller than the mean, a large IPR corresponding to sectorial information in the $2^{nd}$ or $3^{rd}$ largest eigenvalues and the IPR for the small eigenvalues raised.  For the single factor model, we created a \emph{synthetic} correlation matrix using Eqn. (\ref{chol}), with average correlation $(0.204)$ equal to that of the Euro Stoxx 50 over the time period studied.  As shown in Figure~\ref{IPR}(b), the IPR retains some of the features found for empirical data, \cite{Plerou_2000,Noh}, with that corresponding to the largest eigenvector having a much smaller value than the mean.  This corresponds to an eigenvector to which many stocks contribute, (effectively the market eigenvector), \cite{Plerou_2000,Noh}.

In an attempt to include additional empirical features, such as the band of deviating large eigenvalues between the bulk and the largest eigenvalue, we also considered a perturbation, with two groups of stocks having correlation $\rho_{0}-0.15$ and $\rho_{0}+0.15$, and the same average correlation.  In this case, Figure~\ref{IPR}(b), additional features are found, with  larger IPR for both smallest and second largest eigenvalue.  This agrees with \cite{Plerou_2000} where, for empirical data, the group structure resulted in a number of small and large eigenvalues with larger IPR than that of the bulk of eigenvalues.  These large eigenvalues were shown, \cite{Plerou_2000,Conlon_2007}, to be associated with correlation information related to the group structure.

\subsection{Drawdown Analysis}
As described, \cite{Drozdz_2000}, and discussed above, drawdowns (periods of large negative returns) tend to reflect an increase of one eigenstate of the cross-correlation matrix, with the opposite occurring during drawups. In this section, we attempt to characterise the market according to the relative size of both the largest and small eigenvalues using stocks of the Dow Jones Euro Stoxx 50.

Returns, correlation matrix and eigenvalue spectrum time series for overlapping windows of $200$ days were calculated and normalised using the mean and standard deviation over the entire series, (Eqn.~\ref{normalise}).  Representing normalised eigenvalues in terms of standard deviation units (SDU) allows partitioning according to the magnitude of the eigenvalues.  Table~\ref{evalueReturns} shows the average return of the index, (during periods when the largest eigenvalue is $\pm1$ SDU), for both the largest eigenvalue and the average of the normalised $40$ smallest eigenvalues.

The results, Table~\ref{evalueReturns}, demonstrate that when the largest eigenvalue is $>1$ SDU, the average index return over a $200$ day period is found to be $-16.80\%$.  When it is small ($<-1$ SDU), the average index return is $18.46\%$.  Hence, the largest eigenvalue can be used to characterise markets, with the eigenvalue peaking during periods of negative returns (Drawdowns) and bottoming out when the market is rising (Drawup).  For the average of the $40$ smallest eigenvalues, the partition also reflected drawdowns and drawups, but with opposite signs.  This indicates that information about the correlation dynamics of financial time series is visible in both the lower and upper eigenstates, in agreement with the results found by \cite{Schindler_2007,Muller_2005} for both synthetic data and EEG seizure data.

\begin{table*}[htbp!]
	\centering
		\begin{tabular}{c c c c}
				& \emph{Eigenvalues} & \emph{No. Std}  &  \emph{Index Return} \\
		\hline
		\hline
		
				 & &  $<$-1 & 18.46\%  \\[0ex]	
	
			   & \raisebox{1.5ex}{\emph{Large}} & $>$1 & -16.80\%  \\[0ex]
			   
				 & &  $<$-1 & -23.90\% \\[0ex]	

			   & \raisebox{1.5ex}{\emph{Average 40 Smallest}} & 	$>$1 & 19.32\% \\[0ex]					 
								
		\end{tabular}
		\caption{Drawdown/Drawup analysis, average index returns for eigenvalue partitions in SDU.}
			\label{evalueReturns}
\end{table*}

\section{Conclusions}
The correlation structure of multivariate financial time series was studied by investigation of the eigenvalue spectrum of the equal-time cross-correlation matrix.  By filtering the correlation matrix through the use of a sliding window we have examined behaviour of the largest eigenvalue over time.  As shown, Figures (\ref{S&PeigsNorm} -~\ref{SX5E}), the largest eigenvalue moves counter to that of a band of small eigenvalues, due to \textit{eigenvalue repulsion}. A decrease in the largest eigenvalue, with a corresponding increase in the small eigenvalues, corresponds to a redistribution of the correlation structure across more dimensions of the vector space spanned by the correlation matrix.  Hence, additional eigenvalues are needed to explain the correlation structure in the data.  Conversely, when the correlation structure is dominated by a smaller number of factors (eg. the ``single-factor model'' of equity returns), the number of eigenvalues needed to describe the correlation structure in the data is reduced.  In the context of previous work, \cite{Drozdz_2000,Drozdz_2001}, this means that fewer eigenvalues are needed to describe the correlation structure of `drawdowns' than that of `draw-ups'.

By introducing a simple `one-factor model' of the correlation in the system (Section~\ref{corr_model}), we were able to reproduce the main results of the empirical study.  The compensatory dynamics, described, were clearly seen for a correlation matrix with all elements equal to the average of the empirical correlation matrix.  The model was then adapted, by the addition of pertubations to the correlations, with the average correlation remaining unchanged.  This `markets plus sectors' type model was able to reproduce additional features of the empirical correlation matrix with eigenvalues deviating from below and above the bulk. The Inverse Participation Ratio of the ``markets plus sectors'' model was also shown to have group characteristics typically associated with known Industrial Sectors, with a larger than average value for the smallest eigenvalue and for the second largest eigenvalue.  Through a partition of the time-normalised eigenvalues, it was then shown quantitatively that the largest eigenvalue is greatest/smallest during drawdowns/drawups, and vice versa for the small eigenvalues.

Suggested future work includes a more detailed study of the relationship between the direction of the market and magnitude of the eigenvalues of the correlation matrix.  Studying the multiscaled correlation dynamics over \emph{different granularities} may shed some light on the different collective behaviour of traders with different strategies and time horizons.  Additional analysis of high frequency data may also be useful in the characterisation of correlation dynamics, especially prior to market crashes. Study of the possible relationship between the dynamics of the small eigenvalues and additional correlation information which, according to some authors, \cite{Guhr_2003,Burda_2004_a,Burda_2004_b,Malevergne,Kwapien_2006,Muller_2005,Muller_2006_b}, may be hidden in the part of the eigenvalue spectrum normally classifed as noise could be achieved through analysis of the relative dynamics of the small and large eigenvalues at times of extreme volatility, (such as during market crashes).


\begin{thebibliography}{label}
\bibitem{Laloux_1999}
Laloux, L., Cizeau, P., Bouchaud, J., Potters, M.,  \textit{Noise dressing of financial correlation matrices}, Phys. Rev. Lett. 83 (7) (1999) 1467\verb`-`1470.
\bibitem{Plerou_1999}
Plerou, Gopikrishnan, Rosenow, Amaral, Gurh, T, Stanley, H.E., \textit{Universal and non-universal properties of cross-correlations in financial time series}, Phys. Rev. Lett. 83 (1999) 1471\verb`-`1474.
\bibitem{Laloux_2000}
Laloux, L., Cizeau, P., Potters, M., Bouchaud, J., \textit{Random matrix theory and financial correlations}, Int. J. Theoret. \& Appl. Finance 3 (3) (2000) 391\verb`-`397.
\bibitem{Plerou_2000}
Plerou, V., Gopikrishnan, P., Rosenow, B., Amaral, L., Stanley, H.E., \textit{A random matrix approach to cross-correlations in financial data}, Phys. Rev. E 65 (2000) 066126.
\bibitem{Gopikrishnan_2000}
Gopikrishnan, P., Rosenow, B., Plerou, V., Stanley, H.E., \textit{Identifying business sectors from stock price fluctuations}, Phys. Rev. E 64 (2001) 035106R.
\bibitem{Utsuki_2004}
Utsugi, A., Ino, K., Oshikawa, M., \textit{Random matrix theory analysis of cross-correlations in financial markets}, Phys. Rev. E 70 (2004) 026110.
\bibitem{Bouchaud_book}
Bouchaud, J. P., Potters. M, \textit{Theory of Financial Risk and Derivative Pricing},
Cambridge University Press, 2003.
\bibitem{Wilcox_2004}
Wilcox, D., Gebbie, T., \textit{On the analysis of cross-correlations in South African market data}, Physica A 344 (1\verb`-`2) (2004) 294\verb`-`298.
\bibitem{Sharifi_2004}
Sharifi, S., Crane, M., Shamaie, A., Ruskin, H.J., \textit{Random matrix theory for portfolio optimization: a stability approach}, Physica A 335 (3\verb`-`4) (2004) 629\verb`-`643.
\bibitem{Conlon_2007}
Conlon, T., Ruskin, H.J., Crane, M., \textit{Random matrix theory and fund of funds portfolio optimisation}, Physica A 382 (2) (2007) 565\verb`-`576.
\bibitem{Conlon_2008}
Conlon, T., Ruskin, H.J., Crane, M., \textit{Wavelet multiscale analysis for Hedge Funds: Scaling and strategies}, To appear: Physica A (2008), doi:10.1016/j.physa.2008.05.046 
\bibitem{Podobnik_2008}
Podobnik, B., Stanley, H.E., \textit{Detrended cross-correlation analysis: A new method for analysing two non-stationary time series}, Phys. Rev. Lett 100 (2008) 084102.
\bibitem{Schindler_2007}
Schindler, K., Leung, H., Elger, C.E., Lehnertz, K., \textit{Assessing seizure dynamics by analysing the correlation structure of multichannel intracranial EEG}, Brain 130 (2007) 65\verb`-`77.
\bibitem{Schindler_2007_b}
Schindler, K., Elger, C.E., Lehnertz, K., \textit{Increasing synchronization may promote seizure termination: Evidence from status epilepticus}, Clin. Neurophysiol. 118 (9) (2007) 1955\verb`-`1968
\bibitem{Kwapien_2000}
Kwapien, J., Drozdz, S., Ionannides A.A., \textit{Temporal correlations versus noise in the correlation matrix formalism: an example of the brain auditory response}, Phys. Rev. E 62 (2000) 5557
\bibitem{Ying_1966}
Ying, C.C., \textit{Stock market prices and volumes of sales}, Econometrica 34, (1966) 676.
\bibitem{Karpoff_1987}
Karpoff, J.J., \textit{The relationship between price changes and trading volumes: A survey}, Financ. Quant. Anal. 22, (1987) 109.
\bibitem{LeBaron_1999}
LeBaron, B., Arthur, W.B., Palmer, R., \textit{Time series properties of an artificial stock market}, J. Econ. Dyn. Control 23, (1999) 1487.
\bibitem{Guhr_2003}
Guhr, T., K\"{a}lber, B., \textit{A new method to estimate the noise in financial correlation matrices}, J. Phys. A 35 (2003) 3009\verb`-`3032
\bibitem{Burda_2004_a}
Burda, Z., G\"{o}rlich, A., Jarosz, A., Jurkiewicz, J., \textit{Signal and noise in correlation matrices}, Physica A 343 (2004) 295\verb`-`310
\bibitem{Burda_2004_b}
Burda, Z., Jurkiewicz, J., \textit{Signal and noise in financial correlation matrices}, Physica A 344 1\verb`-`1 (2004) 67\verb`-`72
\bibitem{Malevergne}
Malevergne, Y., Sornette, D., \textit{Collective origin of the coexistence of apparent random matrix theory noise and of factors in large sample correlation matrices}, Physica A 331 3\verb`-`4 (2004) 660\verb`-`668
\bibitem{Kwapien_2006}
Kwapien, J., Drozdz, S., Oswiecimka, P.,  \textit{The bulk of the stock market correlation matrix is not pure noise}, Physica A 359 (2006) 589\verb`-`606
\bibitem{Drozdz_2000}
Drozdz, S., Gruemmer, F., Ruf, F., Speth, J., \textit{Dynamics of competition between collectivity and noise in the stock market}, Physica A 287 (2000) 440\verb`-`449
\bibitem{Drozdz_2001}
Drozdz, S., Gruemmer, F., Ruf, F., Speth, J., \textit{Towards identifying the world stock market cross-correlations: DAX versus Dow-Jones}, Physica A 294 (2001) 226
\bibitem{Muller_2005}
M\"{u}ller, M., Baier, G., Galka, A., Stephani, U., Muhle, H., \textit{Detection and characterization of changes of the correlation matrix in multivariate time series}, Phys. Rev. E 71 (2005) 046116
\bibitem{Muller_2006_b}
M\"{u}ller, M., Jim$\acute{e}$nez Y.L., Rummel, C., Baier, G., Galka, A., Stephani, U., Muhle, H., \textit{Localized short-range correlations in the spectrum of the equal-time correlation matrix}, Phys. Rev. E 74 (2006) 041119
\bibitem{Elton}
Elton, E.J., Gruber, M.J., Brown, S.J., Goetzmann, W., \textit{Modern Portfolio Theory and Investment Analysis}, $7^{th}$ edition, Wiley, 2006.
\bibitem{Noh}
Noh, J.D., \textit{A model for stock correlations for stock markets}, Phys. Rev. E 61 (2000) 5981-5982.
\bibitem{Papp}
Papp, G., Pafka, Sz., Nowak, M.A., Kondor, I., \textit{Random matrix filtering in portfolio optimization}, Acta Phys. Pol. B 36 (2005) 2757.
\end{thebibliography}
\end{document}